\documentclass[twocolumn]{aastex62}

\DeclareGraphicsExtensions{.jpg,.pdf,.png,.eps,.ps}

\usepackage{amsmath,amsbsy}
\newcommand{\comment}[1]{{}}
\newcommand{\lcdm}{\ensuremath{\Lambda\mathrm{CDM}}}

 \def\be{\begin{equation}}
\def\ee{\end{equation}}
\def\ba{\begin{eqnarray}}
\def\ea{\end{eqnarray}}
 
 \newcommand{\planck}{\textit{Planck}}

\begin{document}

\title{Cleaning our own Dust: Simulating and Separating Galactic Dust Foregrounds with Neural Networks}
\author{K. Aylor}
\affiliation{Department of Physics, University of California, Davis, CA, USA 95616}
\author{M. Haq}
\affiliation{Department of Mathematics, University of Texas, Dallas, TX, USA 75080}
\author{L. Knox }
\affiliation{Department of Physics, University of California, Davis, CA, USA 95616}
\author{Y. Hezaveh }
\affiliation{D\'epartement de Physique, Universit\'e de Montr\'eal, Montreal, Quebec, Canada H3T 1J4}
\affiliation{Center for Computational Astrophysics, Flatiron Institute, New York, NY, USA 10010}
\author{L. Perreault-Levasseur }
\affiliation{D\'epartement de Physique, Universit\'e de Montr\'eal, Montreal, Quebec, Canada H3T 1J4}
\affiliation{Montreal Institute for Learning Algorithms, Universit\'e de Montr\'eal, Montreal, Quebec, Canada H2S 3H1}
\affiliation{Center for Computational Astrophysics, Flatiron Institute, New York, NY, USA 10010}

 \email{kmaylor@ucdavis.edu}

\begin{abstract}
Separating galactic foreground emission from maps of the cosmic microwave background (CMB), and quantifying the uncertainty in the CMB maps due to errors in foreground separation are important for avoiding biases in scientific conclusions. Our ability to quantify such uncertainty is limited by our lack of a model for the statistical
distribution of the foreground emission. Here we use a Deep Convolutional Generative Adversarial Network (DCGAN) to create an effective non-Gaussian statistical model for intensity of emission by interstellar dust. For training data we use a set of dust maps inferred from observations by the Planck satellite. A DCGAN is uniquely suited for such unsupervised learning tasks as it can learn to model a complex non-Gaussian distribution directly from examples. We then use these simulations to train a second neural network to estimate the underlying CMB signal from dust-contaminated maps. We discuss other potential uses for the trained DCGAN, and the generalization to polarized emission from both dust and synchrotron.
\end{abstract}


\section{Introduction}
    Polarized emission from the interstellar medium of the Milky Way, in the cleanest parts of the sky at the cleanest observing frequencies, is comparable to the cosmic microwave background signal generated by primordial gravitational waves (PGW) if the PGW signal is near the current upper limit. The current upper limit, quantified by the ratio of tensor-to-scalar fluctuation power, $r$, is $r < 0.07$ at 95\% confidence \citep{bicep2keck15}. So-called Stage III CMB experiments, such as the Simons Observatory \cite{SimonsForecast}, and BICEP Array \citep{Cukierman19} combined with SPT-3G \citep{benson14} are designed to have sufficient sensitivity and systematic error control to tighten the 95\% confidence upper limits by a factor of about 20. The Stage IV experiments LiteBIRD and CMB-S4 are targeting upper limits factors of 2 and 5 times more stringent still, respectively. Thus we are rapidly moving into a regime where the foreground contamination is up to two orders of magnitude larger\footnote{This is for fluctuation power. The rms level of contamination in the map is up to one order of magnitude larger than the signal of interest.} than the signal of interest. 

The most exciting possibility is that there will be a detection of PGW, as opposed to improved upper limits. A detection claim would essentially be a claim that there is power remaining in the map that cannot be explained as a residual instrumental systematic or residual foreground emission. Detection therefore requires not only foreground cleaning, but the capability to quantify the probability distribution of residual foreground power. Such capability is hampered by our lack of prior knowledge of the probability distribution of the non-Gaussian and spatially non-isotropic galactic foreground emission. 

In this paper we explore the application of neural networks to the challenges of characterizing non-Gaussian foreground emission and cleaning it from CMB maps. 
Although primarily motivated by the need to clean polarized emission, in this paper we describe our initial studies that are of the intensity of dust emission, rather than its polarization.  The intensity of the CMB is also of cosmological interest and our work may also have applications to the extension of usable regions of the sky to areas of higher galactic emission than is possible with traditional foreground-cleaning methods.

Neural networks are a form of machine learning, also known as deep learning, the development of which was loosely based on how signals are transmitted through a nervous system. In general neural networks approximate a target function as a series of affine and non-linear transformations, the weights of which are updated during training through a process known as backpropagation: the error from a loss function is used to adjust the model weights via stochastic gradient descent or some other optimization algorithm. Over the last decade neural networks have become increasingly popular as a method for performing classification and regression as they have been shown to be universal approximators \citep{csaji01}. In the context of CMB analysis, some recent works have applied neural networks to performing Wiener filtering \citep{Munchmeyer19} and lensing reconstruction \citep{Caldeira18}. In an earlier work, \cite{Auld08}, a network was used to emulate the calculation of CMB angular power spectra.

Developments in modeling via neural networks and the availability of powerful computation resources open up a new approach to conducting cosmological analyses. In particular neural networks can be used to create highly accurate simulations based on a training data set without trying to emulate a particular summary statistic, and can perform map level component separation without relying on a predefined spatial and/or frequency dependence model. The work presented in this paper is meant as a proof of concept and while we focus on intensity maps we plan to extend the work to polarization.  In Section 2 we present a method for developing interstellar dust simulations using a Deep Convolutional Generative Adversarial Network (DCGAN). A DCGAN is a combination of two or more neural networks, which together are capable of learning a generating function; that allows one to sample an unknown distribution (such as the intensity distribution of interstelllar dust). Such a generating function would have many uses including: estimation of statistical properties of foreground residuals, approximation of a likelihood function for a Bayesian sampling approach, or increasing the size of a training set for another deep learning process. We apply the latter in section 3 where we discuss a new approach to separating the CMB from foreground signals using a ResUNet model we have modified to perform regression. Finally we conclude in section 4.

\section{Generative Adversarial Network}

A Generative Adversarial Network (GAN) is a form of unsupervised deep learning that can be used to model a generating function to create samples from a desired distribution \citep{goodfellow14}. 
A GAN consists of two sub-networks; a discriminator and a generator, each with their own sets of weights to be optimized.
The discriminator is optimized to detect samples from $P$, the distribution that we desire to emulate, and the generator is optimized to create samples from $P$.
During the training process the discriminator is shown labeled samples from $P$ and from the generator. As the discriminator improves at detecting samples from $P$ the generator must improve at creating samples belonging to $P$ to minimize its own loss function. Ideally, training proceeds until the generator's output distribution has converged to $P$.
A Deep Convolutional GAN (DCGAN), first presented in \cite{radford15}, is a particular design of GAN where the generator $\rm{G}(z)$ maps $z$, a random vector from $Z$, to $P$ in $\mathbb{R}^{N\times M}$ through a series of upsamplings via strided convolutions. 
The generator then allows one to produce samples from $P$ by sampling $Z$. 

In our case $P$ is the intensity of thermal emission from interstellar dust across the sky.
A DCGAN allows us to generate simulations of dust intensity maps based on the actual measured intensity.
As we only have one sky to measure we are limited in our ability to measure samples from $P$ and instead focus on a subset of $P$, patches of sky with approximately 1\% coverage of the full sky. 
The primary reason for choosing 1\% versus some other size is this coverage reduces the computational power needed to develop a model while still covering angular scales of interest and allowing for the creation of a sufficiently large training set.

\subsection{The Dataset}
We formed our training dataset from the \planck\ 353GHz GNILC intensity dust map \citep{planck16-48}. The map was cut into square patches of approximately 1\% sky coverage using the healpy and HEALPix\footnote[1]{http://healpix.sourceforge.net} package \citep{gorski05,zonca19}.  One can envision our sampling process as shifting the center of a patch at a given longitude and latitude, $(\phi,\theta)$, to  $(\phi+s/cos(\theta),\theta+s)$, where $s$ is the step size,  and selecting a $20^{\rm{o}} \times 20^{\rm{o}}$ region centered on a great circle going through the new center and parallel to the top and bottom edges of the new patch. The factor of $1/cos(\theta)$ is included to make the step in longitude the same anglular separation as the step in latitude. We also exclude the galactic plane by only sampling regions 15 degrees above and below the plane as we are interested in the properties of the dust at high latitudes. 

For $s=5$ degrees we split the full sky map into 1034 smaller maps.  We chose a resolution of $256\times256$ as this allows for easier training than trying to match the \planck\ resolution; with a larger network and more computational power one could simulate maps at a greater resolution.
The average angular size of an individual pixel is less than 5 arcmin. Before training we take the log of each pixel (to reduce the dynamic range and lower the influence of the tails of the distribution) and normalize the entire dataset to the range $[-1,1]$. We note that while we use actual measurements of galactic dust intensity as our training set in this paper, to expand to polarization one would likely have to resort to using simulations as the training set. In such cases the GAN would act as an emulator of the more computationally expensive simulations.

\subsection{DCGAN Architecture}
We base the architecture of our discriminator and generator on the guidelines presented in \cite{radford15} with several notable exceptions.
We replaced all transpose convolution layers in the generator with a bi-linear upsampling followed by a convolutional layer with a stride, or step size, of one unit.
We found this method led to better convergence in the generator by eliminating the checkerboarding artifact that can be found with transpose convolution layers \citep{odena16}.
The generator receives a 64-dimensional vector drawn from $\mathcal{N}(0,1)$ as input, that is then passed through a densely connected layer and reshaped into 512 $16 \times 16$ pixel maps.
This is followed by 4 layers of upsampling and convolution that result in a $256\times256$ pixel map.
After each linear layer in the generator we apply a LeakyReLU  activation \citep{maas13}, with a slope of 0.2 over the negative domain, except in the final layer where we apply a hyperbolic-tangent (tanh) activation.
We also apply Batch Normalization \citep{ioffe15}, with a momentum of 0.9, after each activation layer except the final one. In Table \ref{tab:gan_layers} we list the structure of the generator.

\begin{table}[ht]
\centering

 \begin{tabular}{ c c c c }
 \hline\hline
  Operation & Output & Hyperparameters \rule{0pt}{2.5ex} \\ [0.5ex]
 \hline
  Linear & $16\times 16\times 512$  & \rule{0pt}{3.0ex} \\
  Leaky ReLU & $16\times 16\times 512$ & $\alpha=0.2$ \\
  Batch Normalization & $16\times 16\times 512$ & $\rm{momentum} = 0.9$ \\  
  Up Sampling & $32\times 32\times 256$ & bi-linear \\ 
  Convolution & $32\times 32\times 256$ &  \\
  Leaky ReLU & $32\times 32\times 256$ & $\alpha=0.2$ \\
  Batch Normalization & $32\times 32\times 256$ & $\rm{momentum} = 0.9$ \\  
  Up Sampling & $64\times 64\times 128$ & bi-linear \\ 
  Convolution & $64\times 64\times 128$ &  \\
  Leaky ReLU & $64\times 64\times 128$ & $\alpha=0.2$ \\ 
  Batch Normalization & $64\times 64\times 128$ & $\rm{momentum} = 0.9$ \\  
  Up Sampling & $128\times 128\times 64$ & bi-linear \\ 
  Convolution & $128\times 128\times 64$ &  \\
  Leaky ReLU & $128\times 128\times 64$ & $\alpha=0.2$ \\ 
  Batch Normalization & $128\times 128\times 64$ & $\rm{momentum} = 0.9$ \\  
  Up Sampling & $256\times 256\times 1$ & bi-linear \\ 
  Convolution & $256\times 256\times 1$ &  \\
  Tanh & $256\times 256\times 1$ & $\alpha=0.2$ \\ [1ex]
 \hline
\end{tabular}
\caption{The output structure and relevant hyperparameters for each layer in the generator.}
\label{tab:gan_layers}
\end{table}

The architecture for our discriminator model is where we deviate from the standard DCGAN the most. Instead of using just a single discriminator we employ two. 
One discriminator receives a map as input and the other receives the fractional difference of the angular power spectrum with respect to the mean power of the \planck\ maps ($C_{\ell}/\tilde{C}^{\planck}_{\ell}-1$) as input; we refer to them as the map and power discriminators respectively. 

For the map discriminator we use the same number of feature maps as in the generator. The upsampling and convolution steps in the generator are replaced by a convolution in the discriminator with a stride of 2 units. After each convolution we again apply a LeakyReLU activation and Batch Normalization with the same slope and momentum as in the generator. The feature maps are then flattened into a 1-dimensional vector and passed through a densely connected layer with a sigmoid activation function. 

The architecture of the power discriminator is largely the same as the map discriminator, except convolutions in two dimensions are replaced by convolutions over a single dimension and the overall size of the power discriminator is smaller. 
The power discriminator only has three convolution layers and the number of features increases from 1 to 256 in multiples of 64.

In the production of this work we tested various network architectures. We do not claim to have found an optimal network but simply one that performs better than alternatives we have tried. Since the development of the DCGAN, other architectures have risen in popularity, in particular architectures employing the Wasserstein loss function, such as WGAN, WGAN with gradient penalty, and CTGAN \citep{Arjovsky2017,Gulrajani17,Wei18}. We tested these networks but in all of our trials we found the generator results to be significantly inferior to those produced by our best DCGAN, even before we implemented the second discriminator.

\subsection{Training}
We used a binary cross entropy loss function and the Adam optimizer \citep{kingma15} with the learning rate set to $2\times10^{-4}$ and the first and second momentum parameters to 0.5 and 0.999 respectively for each discriminator and the final loss used to update the network is the sum of the map and power discriminator losses. Training is done in batches of 32 maps. First the discriminator is trained on 32 real images and then 32 fake images (produced by the current state of the generator).
The images' labels are also swapped with 1\% probability, to avoid the discriminator overpowering the generator.
Next the generator is given 32 random noise vectors with dimension 64 drawn from a normal distribution and the output from the generator is passed to the discriminators to calculate the loss. 

The statistic of greatest interest to us is the distribution of the power spectrum; we base our stopping criteria for training on this. After every 100 training steps we generate 1034 simulations, restore the original range, and calculate the power spectrum for each map. We then calculate the Fr\'echet distance ($d_F$) between the real and simulated distributions of the log of the power spectrum\footnote{We apply the $\rm{log}_{10}$ function to the power spectra in order to work with distributions that are less skewed and closer to normal distributions.}, 
which for multivariate normal distributions takes the following form,
\begin{equation}
    d_F = |\mu_r - \mu_s|^2 +  tr(\Sigma_r + \Sigma_s - 2(\Sigma_r\Sigma_s)^{\frac{1}{2}}).
\end{equation}
\noindent In the above equation $\mu_i$ and $\Sigma_i$ are the mean and covariance of either the real ($r$) or simulated ($s$) power spectrum. After training for 50,000 steps we take the GAN state with the minimum $d_F$ and train it for another 5000 steps, this time calculating $d_F$ after every step. We then take the state with the minimum $d_F$ as our best-fit model. We note this choice of metric is insensitive to the tails of the distributions but despite the training distribution being non-Gaussian we found this metric to be computationally efficient and lower values of $d_F$ correlated with improved results.
Training was done on the Extreme Science and Engineering Discovery Environment (XSEDE) Comet GPU resource \citep{xsede}.

\subsection{Results}

\begin{figure*}[!tbh]
\includegraphics[width = \textwidth]{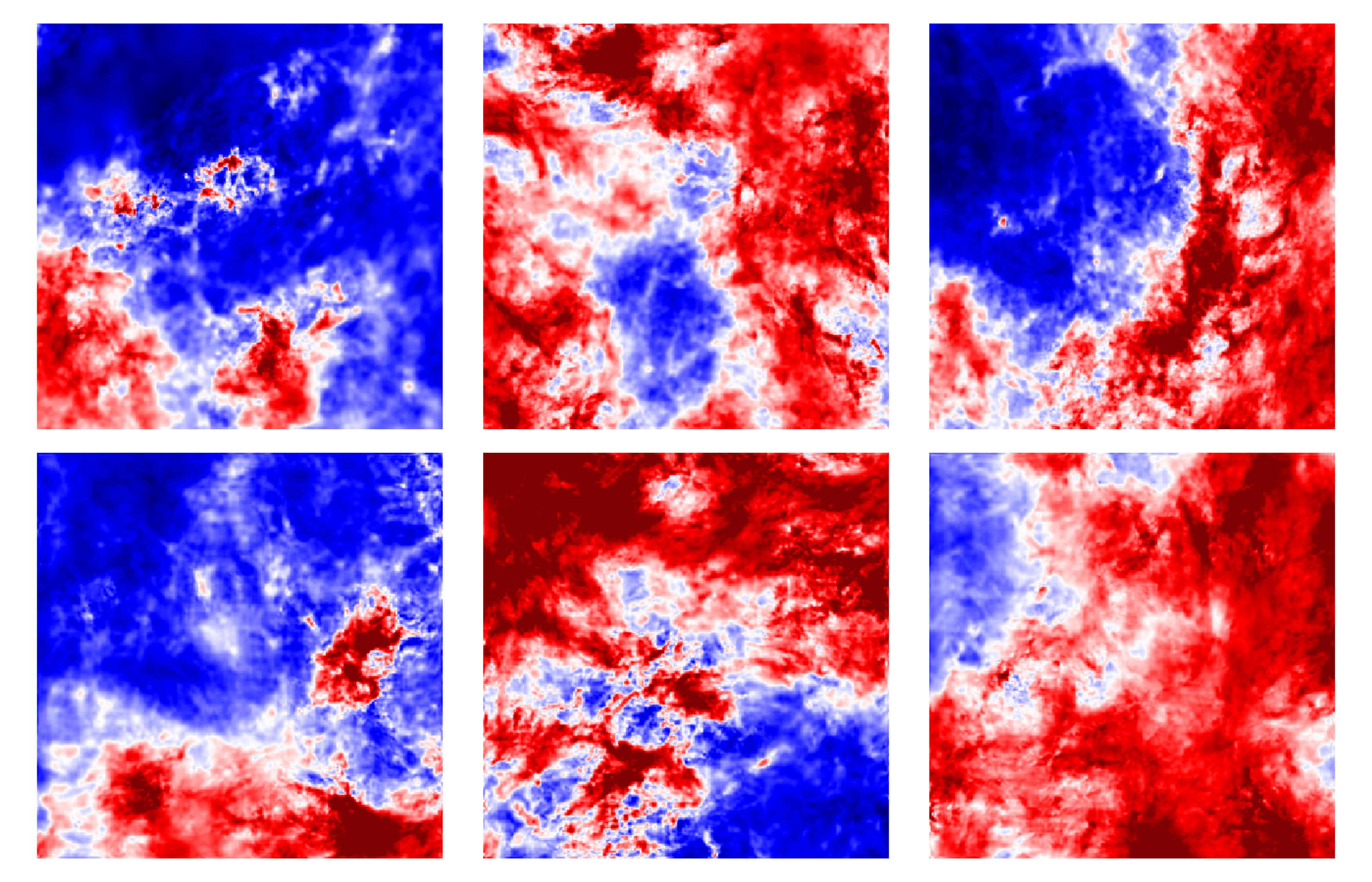}
\caption{A random selection of images from the training set (top row) and from our GAN (bottom row).}
\label{fig:random_samples}
\end{figure*}

Determining the quality of samples produced by the generator of a GAN is a current area of research and several methods have been proposed. We choose to follow the methodology presented in
 \cite{mustafa17} where quality is determined by the GAN's ability to replicate relevant summary statistics. The three statistics we compare for the real and simulated datasets are the pixel intensity, power spectra, and Minkowski functional distributions. The first two statistics capture the one- and two-point function information while the third is sensitive to  higher-order correlations, that are of interest since the distribution of dust intensity is highly non-Gaussian. 

\begin{figure}
\includegraphics[width = .5\textwidth]{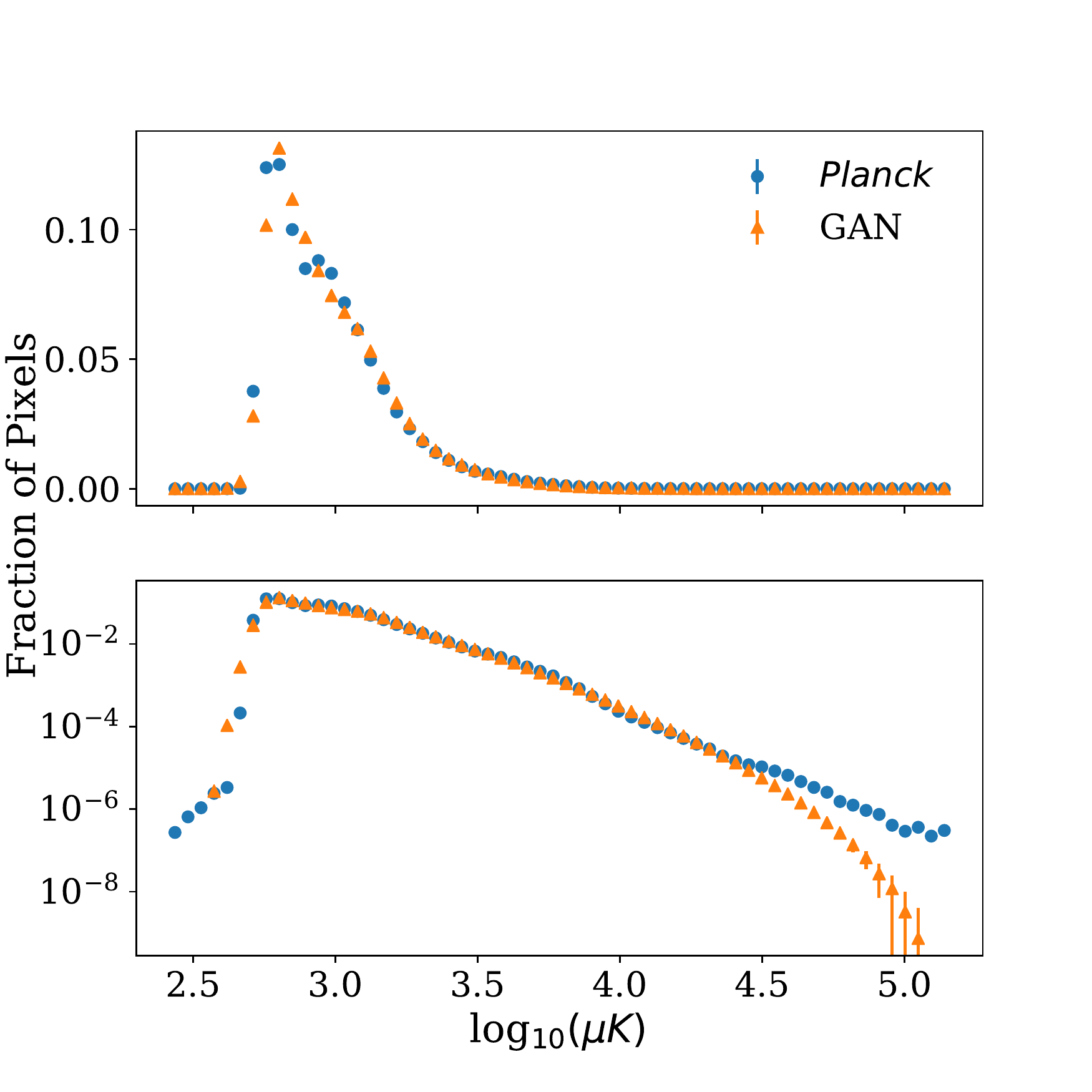}
\caption{The pixel intensity distribution for the \planck\ and generated maps. In the bottom panel we show the distribution on a log scale to highlight the differences between the simulations and the real data at the tails.}
\label{fig:pix_int}
\end{figure}

We begin by showing a random selection of images from the training set and generator in Figure \ref{fig:random_samples}. The generated images appear to have similar features to the training set and no obvious visual artifacts.
In Figure \ref{fig:pix_int} we show the distribution of pixel intensities over the entire set of
real maps and an equal number of generated maps. From Figure \ref{fig:pix_int} we see the
GAN does not produce the same pixel intensity distribution as the training data but does capture the bulk mass with an average intersection of $94\%$ taken over 1000 bootstrapped samples. The intersection of two histograms with equal binning and number of samples is defined as $\Sigma_{i} min(a_i,b_i)/ (\Sigma_{i} a_i)$, where $a_i$ and $b_i$ are the $i^{\rm{th}}$ bins of the two histograms. 
The GAN does not capture the full range of intensities found in the real distribution and also fails to replicate some of the more subtle features around the peak of the real intensity distribution. The behaviour at the tails is unsurprising as a generator
will have more difficulty learning these regions due to the low rates they are seen during the training. The discrepancies near the peak may be the result of the distribution being too complicated for the GAN to learn as the real distribution is somewhat bi-modal.

Our primary interest in creating these simulations of the dust intensity is to learn and replicate the distribution of the angular power spectrum. 
The power spectrum of an intensity map is the variance in the intensity at different scales; it is the most informative statistic of the CMB and measurements of it have resulted in the tightest constraints on cosmological models. 
Measurements of the CMB power spectrum are contaminated by dust and therefore it is necessary to model the power spectrum of the dust to separate the two signals.

\begin{figure*}[!tbh]
\includegraphics[width = \textwidth]{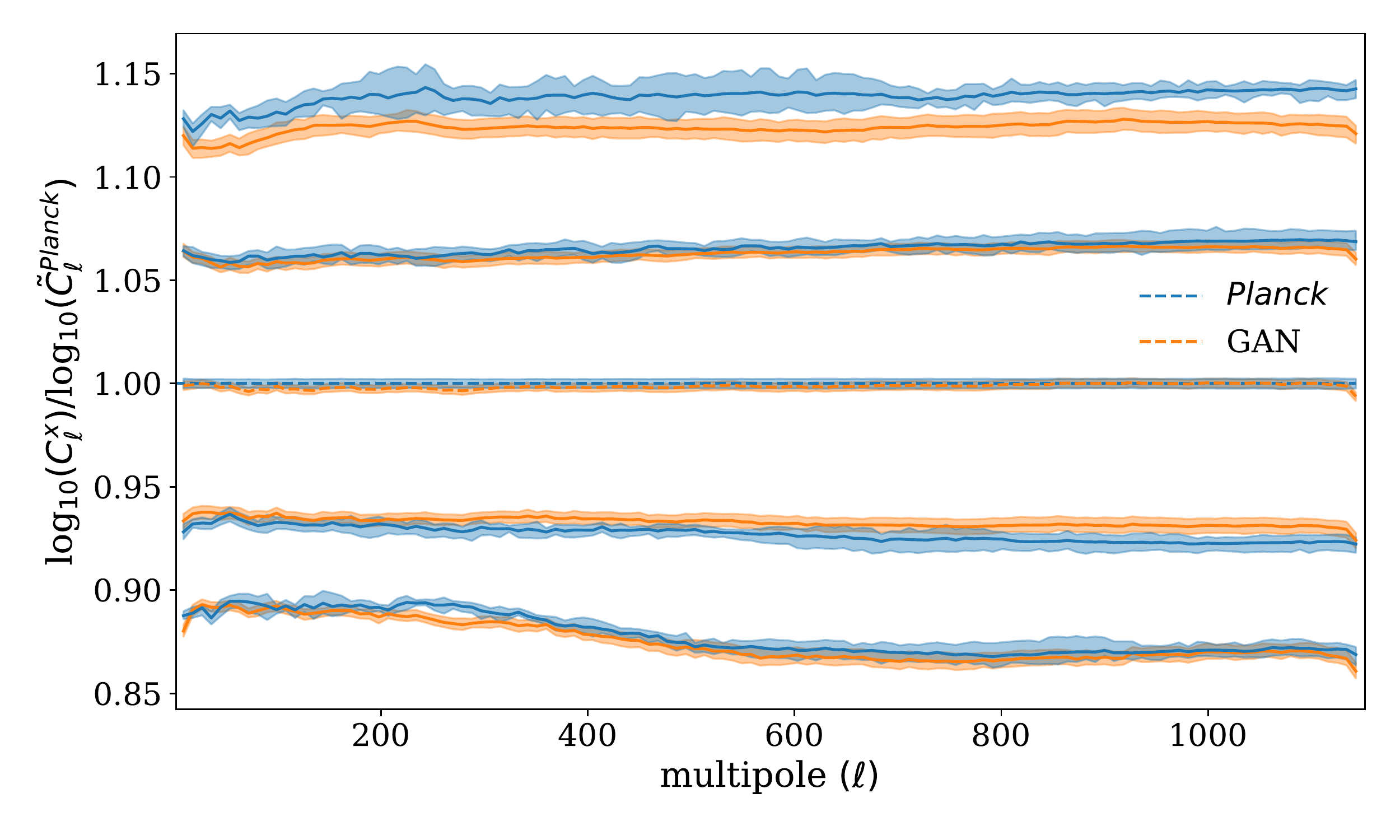}
\caption{The mean (dashed lines in center), 68\%, and 95\% central intervals of the GAN (orange) and \planck\ (blue) log power spectra distributions centered on the \planck\ mean. The errorbar for each statistic was obtained through bootstrapping the \planck\ dust intensity maps or the GAN. The GAN is not capable of emulating the 95\% intervals of the \planck\ distribution as it does not produce the same dynamic range found on the upper end of the \planck\ distribution. The largest discrepancies between the two distributions at the lower end are found between $200 <\ell < 400$.}
\label{fig:power}
\end{figure*}

\begin{figure*}[!tbh]
\includegraphics[width = \textwidth]{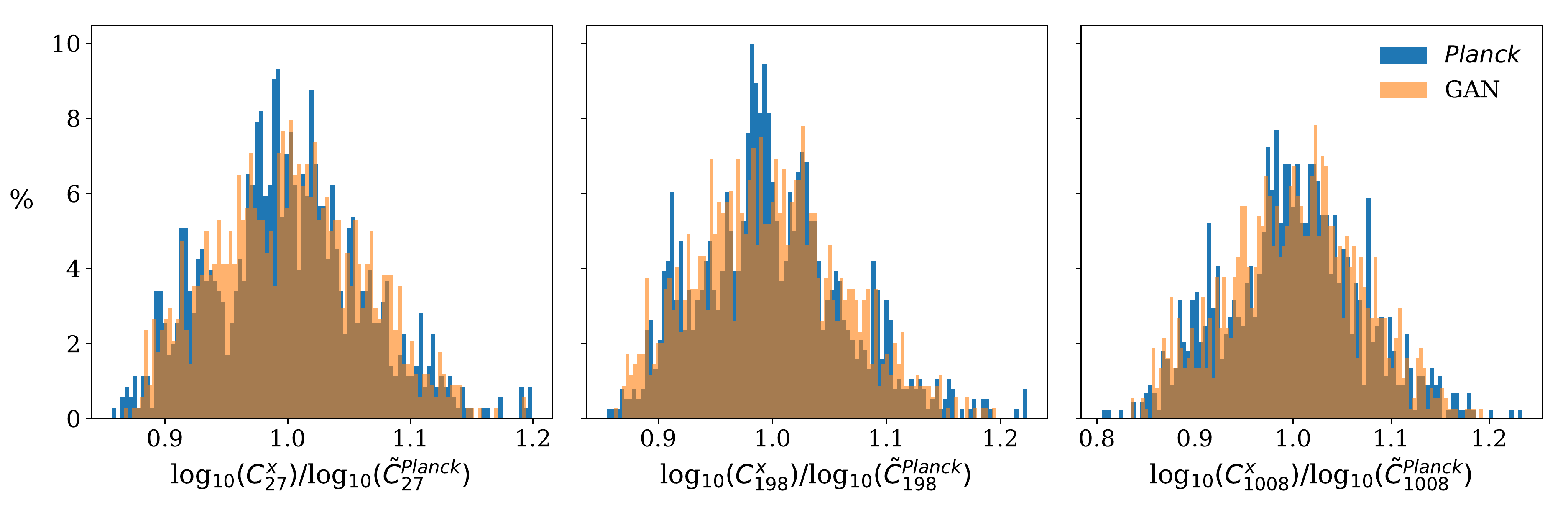}
\caption{The distributions of the $C_{\ell}$ for three separate bins at $\ell=27, 198, 1008$.}
\label{fig:powerhist}
\end{figure*}

The angular extent of our maps is sufficiently small that a flat-sky approximation is sufficiently accurate. We therefore calculate power spectra from 2-D Fourier modes instead of spherical harmonics. 
In Figure \ref{fig:power} we show the mean, 68\%, and 95\% intervals for the real and generated distributions of the log of the power spectrum. For all plots involving power spectra in this paper, each bin has a width of $\Delta\ell = 9$.
To obtain errors on the presented statistics we bootstrap the real and generated distributions by drawing 1000 samples with each sample being the size of the real dataset (1034 maps). 
Just as with the pixel intensity distribution our GAN has captured the majority of the variation found in the real data set but fails to capture the full range. 
A large portion of the difference in the upper 95\% intervals can be attributed to 6 maps in the real data set that have significantly greater power than what is found in the remainder of the data set. 
Evidence for this can be seen in Figure \ref{fig:powerhist} where we show the distribution of power for three scales chosen at random. 
These highly contaminated maps come from a region of the sky just outside of the $30^{\rm{o}}$ band excluded from the creation of the training data set.

Each panel in Figure \ref{fig:powerhist} indicates the real distribution of power has a heavy tail towards greater power at all scales, that the GAN does not capture well. 
We found that by increasing the variance of the normal distribution used to sample the latent space the GAN will produce more samples with power spectra similar to those found at the higher end of the real distribution. 
The inability to recover the tails is therefore not due to the GAN being unable to create maps with greater power.
We also note that there are discrepancies between the two power distributions at the lower end predominantly for $200 <\ell < 400$.
The discrepancies at the tails between the GAN distribution and the real distribution may not be due to just the infrequency at which the samples are seen during training but may also be connected to the non-trivial mapping from the latent space.
Better results could potentially be obtained by sampling the latent space from a distribution that better matches the distribution to be emulated instead of a Gaussian.
We leave the exploration of this issue to future work.

We also note that in terms of failing to recover the underlying distribution of power the GAN has done so in what could be considered the best possible way. 
When making measurements of the CMB it is desirable to avoid measuring highly contaminated regions and therefore it is not necessary for the GAN to be able to produce the full upper range of the dust intensity power spectrum distribution. 
Also on the lower end it is better to produce a greater level of variation than too little as to be sure to capture all of the possible types of contamination likely to be measured in an actual experiment. 
The GAN's inability to properly recover the tails of the real power distribution indicates that if one were to extend this work to polarization and the detection of primordial gravitational waves it would be better to train a new GAN on the least dynamic range possible that can contain the truth as a means of simplifying the distribution to be learned.

\begin{figure*}[!tbh]
\includegraphics[width = \textwidth]{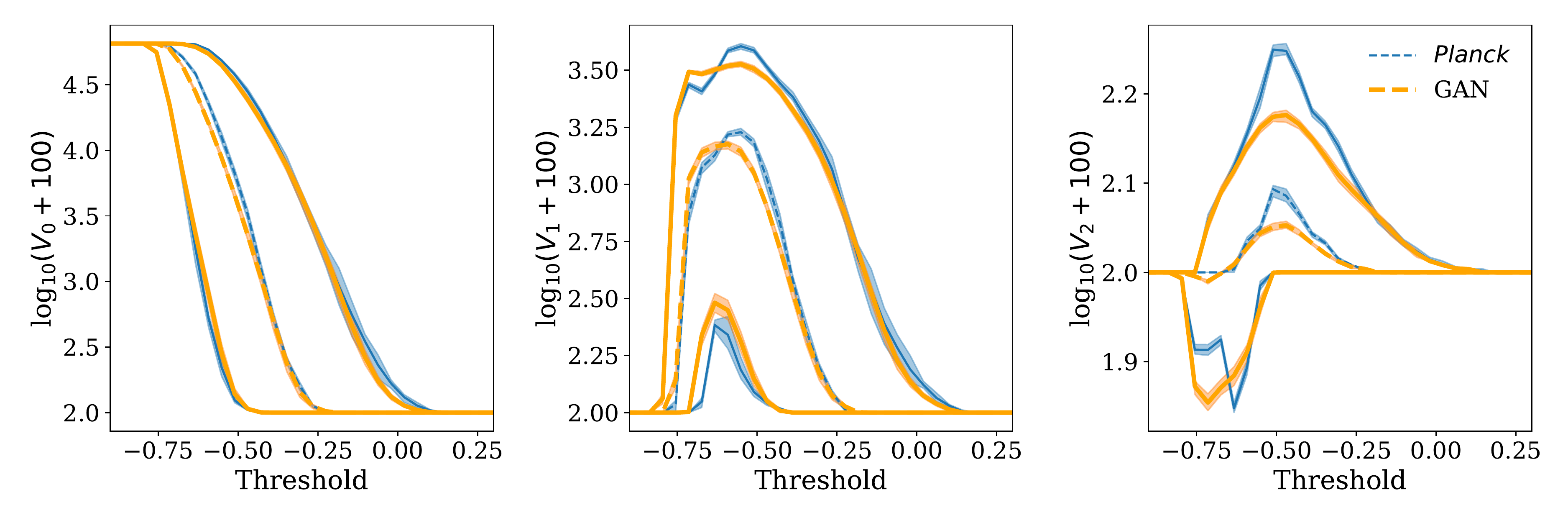}
\caption{The median (dashed lines in center), 68\%, and 95\% central intervals of the GAN (orange) and \planck\ (blue) Minkowski functional distributions for the real (blue) and generated (orange) intensity maps. 
The errorbar for each statistic was obtained through bootstrapping the \planck\ dust intensity maps or the GAN. The map sets were normalized to the range of [-1,1].}
\label{fig:mink}
\end{figure*}

The dust maps contain non-Gaussian information that is not captured by the distribution of pixel intensities  and only somewhat captured by the distribution of the power spectra. 
In order to compare the real and generated sets in a manner sensitive to the non-Gaussian information we use the Minkowski functionals $V_0$, $V_1$, and $V_2$, that respectively measure the area of the foreground, the perimeter of the foreground, and the connectivity of the foreground for various thresholds. 
In Figure \ref{fig:mink}  we show the functionals evaluated at 50 different threshold values after normalizing the map sets to the range [-1,1]. 
Errorbars are again obtained through bootstrapping the real and simulated data. 
It is here that we find the greatest level of disagreement between the two data sets, especially in the  $V_2$ functional distributions. 
From Figure \ref{fig:mink} it is clear the GAN has not captured the full amount of variation found in the training data and in particular struggles the most where the median values of $V_1$, and $V_2$ are largest. 
For all three functionals the GAN fails to recover the median for some of the threshold values. This is another indicator that the GAN struggles to capture the non-Gaussian nature of the real data set.

Generally validation of a neural network's predictions or outputs is done against a subset of data left out of the training process to test if a model has overfit the data or generalized well. 
However we do not follow this practice for two reasons. 
First, we only have 1034 images in our data set; splitting this into a training and validation set would result in few samples for either set. 
Also, since we are not working with a classification or regression problem we are not concerned about generalization, as we are trying to produce samples from the same distribution the training set was drawn from. 
For a large enough data set the summary statistics for training and validation sets would be the same and fitting the training set well would automatically imply the validation set is also well fit. 
Therefore we choose to maximize our training set and do not create a separate validation set. 
We are then left with the task of showing the generated samples are not simply copies of the training data.
This can be done by exploring the latent space (the distribution from which inputs to the generator are drawn) for any hard transitions. 
We test this via the power spectra and find that drawing many samples does fill the range of the power distribution without significant gaps and even fills some regions that are uncovered by the training data.

Finally we argue that the best test of the quality of our simulations is determined by how well they perform their intended purpose. 
In the next section we discuss how these simulations may be used to train a second neural network to separate the dust signal from the CMB at the map level. 
We train this second network only on the simulations from our GAN and validate and test on real data. 
After training we find this second network is able to separate the two components to a high degree of precision in the test set and conclude that while our GAN has room for improvement, that we believe could be achieved with a more optimal architecture or a better choice of distribution for sampling the latent space, it does indeed produce simulations of dust intensity maps from a similar distribution to that of the true underlying one.

\section{ResUNet for Component Separation}
Developing a methodology for separating foreground components from the CMB using neural networks has been the primary motivation for this work. In this section we discuss a type of neural network, a Bayesian ResUNet, that may be used for such a task and apply it to the separation of the CMB and galactic dust foregrounds. We begin with a brief summary of the ResUNet architecture and how to obtain uncertainties from it. Then we conclude with the results of our work.

\subsection{ResUNet}
A ResUNet is an network architecture based on another network called a UNet and we begin by discussing the progenitor.
The UNet architecture was first presented in \cite{Ronneberger15} as a means for 
segmenting biomedical images into different classes.  A UNet contains an encoding 
path and a decoding path. The encoding path receives an image as input and 
through a series of convolutions downsamples the image into a compressed representation. The decoding path takes the compressed representation through a series of convolution and upsampling layers and builds the target image. Ideally the compressed representation learned by the encoding path retains only the most relevant information for constructing the target. If the input to the encoder is a noisy image and the target is a cleaned version of the image a fully optimized network will drop the information related to the noise and the decoder will reconstruct the desired component.  

The encoding and decoding paths can also be broken down into blocks that perform operations at a given scale and then re-scale (downsampling or upsampling) the input before passing it on to the next block. 
In a UNet encoding blocks acting on a particular scale also pass their output to the decoding block of the same scale where the encoding output is concatenated with the up-sampled output from an earlier block in the decoding path. 
These extra connections allow the network to focus on extracting the most important information at each scale and allow easier flow of gradients during training; often the gradient can shrink significantly before reaching earlier layers in a network without these kinds of connections, slowing down learning. 

We will show that a modified version of the original UNet, known as a ResUNet is well suited to the task of removing foregrounds from images of the CMB.  
A ResUNet was first presented in \cite{Foivos19} and recently applied to performing CMB lensing reconstruction in \cite{Caldeira18}. 
The main difference between a ResUNet and a UNet are residual connections from the beginning to the end of each downsampling or upsampling block. 
A residual connection sends the input of a block through an additional linear layer and adds the result to the output of the same block.
In our case the linear layer is a convolution that transforms the input to the same shape as the output for a given block. 
These residual connections act similarly to the connections between the downsampling and upsampling blocks and allow for better flow of gradients. 
They also potentially simplify the function the network needs to learn. 
For this work we began using a standard UNet architecture but found training proceeded more rapidly and we obtained better results with a ResUNet based architecture.

\subsubsection{Network Uncertainties}
The networks we have described so far in this paper are deterministic, for a given input you will always receive a particular output. For many tasks, including the removal of foregrounds, it is necessary to have a measurement of uncertainty to the degree with which the task has been completed. 
We cannot create a network that can separate the CMB and galactic dust foregrounds with perfect accuracy  due to the stochastic nature of the data and limitations of the network and we need to quantify the level of uncertainty in a prediction. 

Bayesian Neural Networks are a method through which we may extract uncertainties of a prediction by specifying the weights of a network with probabilistic distributions. 
During training the network learns the best distribution to draw weights from instead of learning an immutable value. 
The true posterior of plausible weights usually cannot be evaluated analytically and is replaced with a variational distribution that has an analytic form.
During training the parameters of these analytical distributions are optimized so that the distance between the  variational distributions and the true posteriors is minimized.
The choice of variational distribution is important not only in terms of achieving good results but also for computational efficiency i.e. using a Gaussian distribution for a given network effectively doubles the number of parameters that need to be learned while a Bernoulli distribution does not increase the number of parameters. 

In \cite{Gal15} it was shown that a common method for regularization in neural networks known as Dropout can be recast as an approximation to a Gaussian process. 
Dropout was first presented in \cite{srivastava14a} and involves randomly setting a portion of the inputs to a layer to zero with some predetermined probability. 
We refer the reader to \cite{Gal15}, \cite{Gal16}, and \cite{Kendall17} for discussion of estimating uncertainties with neural networks. 
This process results in the network learning a distribution of possible functions conditional on the training data. 
When one wants to make a prediction on a new input one can treat a single pass of the input through the network with dropout on as sampling from the learned posterior. 
Then to calculate the mean or any other relevant statistic of this posterior one can simply perform a Monte Carlo by passing the new input through the network many times.

\subsubsection{Architecture}
\begin{table}[ht]
\centering

\begin{tabular}{c c c }
 \hline\hline
  Layer & Operation & Output\rule{0pt}{2.5ex} \\ [0.5ex]
 \hline
  $[1]$ & Input & $128\times 128\times 32$ \rule{0pt}{3.0ex} \\
  $[2]$ & Batch Normalization([1]) & $128\times 128\times 32$  \\
  $[3]$ & ReLU([2]) & $128\times 128\times 32$  \\
  $[4]$ & Convolution([3]) & $64\times 64\times 64$ \\
  $[5]$ & Dropout([4]) & $64\times 64\times 64$ \\
  $[6]$ & Batch Normalization([5]) & $64\times 64\times 64$ \\
  $[7]$ & ReLU([6]) & $64\times 64\times 64$ \\  
  $[8]$ & Convolution([7]) & $64\times 64\times 64$ \\
  $[9]$ & Dropout([8]) & $64\times 64\times 64$ \\
  $[10]$ & Convolution([1]) & $64\times 64\times 64$ \\
  $[11]$ & Add([9],[10]) & $64\times 64\times 64$ \\ [1ex]
 \hline
\end{tabular}
\caption{The architecture for a generic encoding block. Our ResUNet uses eight of these blocks on the encoding path. The first encoding block excludes the the first Batch Normalization and ReLU layers; the final encoding block excludes the last two layers of convolution and addition (the residual connection). The final layer of each encoding block is concatenated with the input to the decoding block operating on the same scale.}
\label{tab:encoding_block}
\end{table}

To turn a ResUNet into a Bayesian ResUNet we simply need to add a Dropout layer after every convolution layer in the network.
For a typical encoding block we perform three convolutions with a Dropout layer immediately after each convolution (except after the convolution in the residual connection). All convolutions are performed with 3x3 kernels.
The first convolution halves the resolution of the input and if the number of features is to be increased it is also done here. 
We also insert a Batch Normalization and ReLU layer after the first Dropout layer and at the beginning of every encoding block except the first one. 
In the final layer of an encoding block the residual connection is added to the output of the last dropout layer and this sum is passed to the next encoding block and the corresponding decoding block. 
We use a total of eight encoding blocks. 
Starting with the first block, every other encoding block doubles the number of features and increases the dropout rate (both Dropout layers in each block use the same dropout rate). 
The dropout rates increase from 0.05, to 0.10, to 0.20, and to 0.30. 
In Table \ref{tab:encoding_block} we describe an encoding block in greater detail. 

On the decoding path in each block the feature map from the previous block is upsampled with bi-linear interpolation and is then concatenated with the output from the encoding block of the same scale. 
The remaining layers of the decoding block are the same as the encoding block except the order of the dropout rates is reversed and the number of features are halved every other block.
The final layer of the decoding path of our network is a convolution with a kernel size of one pixel that reduces its input to a single channel image. The resolution of this image is the same as the initial input to the network.

Finally, since we are primarily interested in recovering the power spectrum we added one final layer that calculates the angular power spectrum (multiplied by $\ell(\ell+1)$) of the decoding path output and have the network predict both the cleaned CMB map and the corresponding power spectrum. Adding the power spectrum calculation of the cleaned map to the network and including the output power in the loss function naturally lead to better predictions of the power spectrum compared to only having the network predict the cleaned map. By only having the network predict the cleaned map the predictions were allowed to vary from the truth at the map level in any random manner and we found this produced maps with highly correlated noise at the power spectrum level. Making the network predict the map and the power spectrum constrained the way the predictions were allowed to vary from the truth at the map level and produced maps with less correlated noise.

\subsection{Dataset}
We create our training set from a combination of our galactic dust simulations described in Section 3 and CMB maps generated with $\it{healpy}$. 
The CMB maps are realizations drawn from the 2015 \planck\ TT lowTEB \lcdm\ parameter posterior. 
Each foreground simulation is combined with a random realization of the CMB for a total of 48,000 $20^{\rm{o}} \times 20^{\rm{o}}$ maps. 
For our validation and testing sets we combine each map in our set of \planck\ dust maps with a random realization of the CMB and split this set in two. 
The validation and testing data sets each contain 517 images. 
Finally for each set of data we also simulate maps measured at 143GHz and 545GHz using the brightness model for intensity dust in \cite{planck15-10} that assumes the relationship between the intensity at two different frequencies does not have any spatial dependence and is simply a scaling from one map to the other.

\subsection{Training}
Training a ResUNet is a simpler task than training a GAN. The input to the ResUNet is a three channel image with each channel representing a 143GHz, a 353GHz, and a 545GHz map of the CMB and galactic dust. The target output is the corresponding uncontaminated CMB map and the corresponding uncontaminated power spectrum. For a loss function we simply use the sum of the mean square error for the map and the power spectrum estimates.
We use the Adamax optimizer with a learning rate of 0.002 and set the first and second momentum parameters to $0.9$ and $0.999$ respectively. The ResUNet is trained in batches of 32 with early stopping and a patience level of 10 (the network is trained until the validation loss does not decrease for 10 consecutive epochs). 

\subsection{Results}

\begin{figure*}[!tbh]
\includegraphics[width = \textwidth]{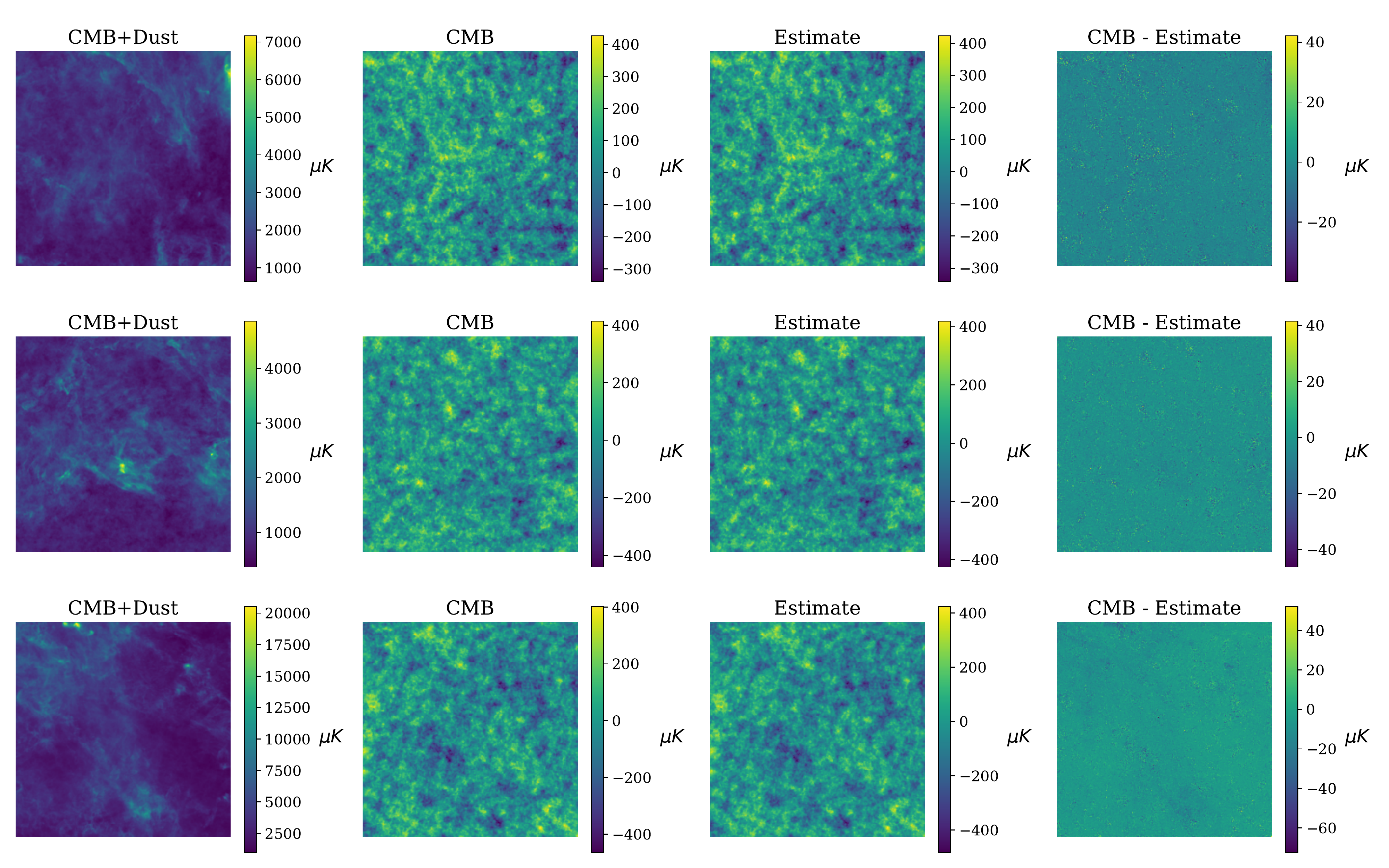}
\caption{The first column shows three CMB maps with varying amounts of galatic dust contamination at 353 GHz (one of the channels used as input into the ResUNet). 
The second column shows the underlying CMB and the third column shows the estimate from the ResUNet. The final column shows the residuals between the underlying CMB and the estimate.}
\label{fig:cleaning}
\end{figure*}

\begin{figure*}[!tbh]
\includegraphics[width = \textwidth]{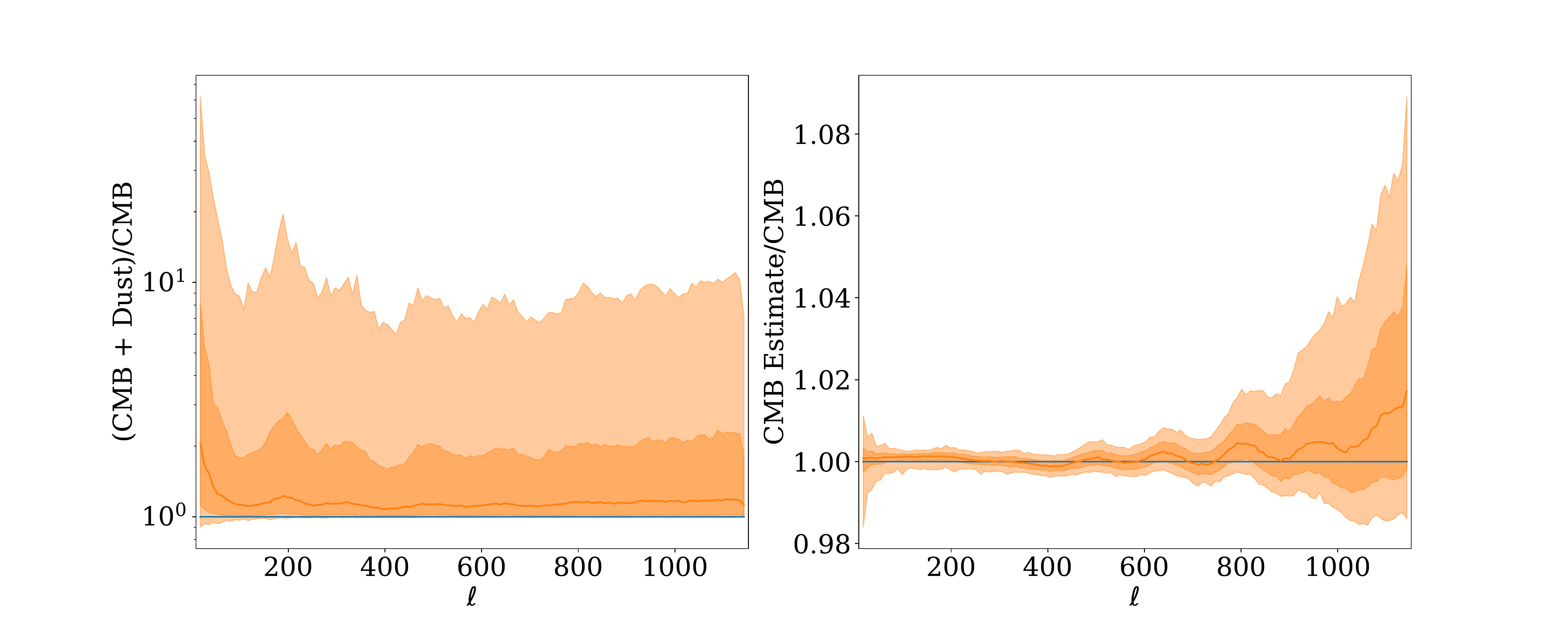}
\caption{\textbf{Left}: The Power spectrum distribution of the CMB maps contaminated with galactic dust relative to the CMB only power spectrum distribution mean at 353GHz.  \textbf{Right}: The distribution of CMB estimates from the ResUNet relative to the CMB only power spectrum distribution mean. In both panels the solid curves are the medians and the shaded regions are the 68\% (darker) and 95\% (lighter) intervals. For the approximate range of $\ell < 700$ the error on the recovered power spectrum is less than 1\% of the underlying CMB signal.}
\label{fig:cleaning_power}
\end{figure*}

In this section we present the degree to which our best ResUNet is able to remove the dust foregrounds from our test data set. 
To make an estimate of the underlying CMB map and power spectrum we pass an image from our test set through the network 1000 times and calculate the mean of the outputs.
In Figure \ref{fig:cleaning} we show the input image, the underlying CMB, the predicted mean, and the residual of the CMB and the prediction for three random images from the test set.  Upon visual inspection it is difficult to detect any difference between the underlying CMB and the prediction. The residual map shows most of the errors are made at small scales with little large scale structure visible. 

We discussed in Section 3 that the most important statistic of the CMB is the power spectrum. Therefore we test our ResUNet's ability to recover the power spectrum of the CMB. In the left panel of Figure \ref{fig:cleaning_power} we show the distribution of the ratio of the contaminated power spectrum to the underlying CMB power spectrum for the entire test set. In the right panel we show the distribution of the ratio of the ResUNet estimate to the underlying CMB power spectrum. We see the ResUNet is able to clean the majority of the test set to less than a 1\% error for approximately $\ell < 700$. Instead of estimating the errors from the posterior of the ResUNet we have taken a Frequentist approach for reasons we will discuss in the next section.

\subsubsection{Accuracy test of Uncertainties}

To evaluate the accuracy of the uncertainties we follow the procedure in \cite{levasseur17}. 
If the posterior distribution predicted by the ResUNet accurately reflects the uncertainties in the cleaning procedure then a confidence interval of a given percent should contain the true value for an equal percent of the test data. 
We can define the coverage probability as the fraction of test samples where the true value lies in a particular confidence interval. 
For an unbiased estimate of a 68\% confidence interval we should find a coverage probability of 68\%. 
We find the average coverage probability over all multipoles to be 93\% for the 68\% confidence interval indicating the network is overly conservative in the estimation of the errors for most multipoles. 
The coverage probability only comes close to achieving desirable levels at the largest and smallest multipoles for our best network. 
We tried many different choices for hyperparameters in order to achieve better coverage probabilities but we found any particular choice lead to errorbars that were either far too conservative or not conservative enough. 
The errorbars predicted by the network are highly dependent on not only the dropout rate but also the number of features in each layer. 
Fine tuning these parameters are beyond the scope of this work and a better understanding of the errors predicted by a Bayesian ResUNet in the context of foreground removal is left to future work.

\section{Summary and Conclusions}
Our work here is motivated by the challenge of detecting or limiting the contributions from tensor perturbations to degree-scale polarization of the CMB in the presence of galactic emission. Here we have conducted a preliminary study, focused on temperature (intensity) rather than polarization, of the effectiveness of neural networks for simulating foreground emission and cleaning foreground emission from measurements of the CMB.

We first showed how a GAN may be used to create simulations of foregrounds from a relatively small training set. 
Our GAN was trained on measurements of the interstellar dust intensity made by the \planck\ satellite at 353GHz. 
From this single map we created approximately 1000 maps with $1\%$ sky coverage.
After exploring a wide range of GAN architectures we found the best results came from a modified version of a DCGAN with two discriminators, one acting at the map level and the other acting at the power level.

Our GAN was able to produce new images that looked to the eye to be similar to real dust maps, and that captured the majority of the variation found in the summary statistics of our training set. Overall we view this initial study as sufficiently successful to motivate training of a GAN to simulate polarized emissions.

Our future work on polarized emission will be informed by some of the shortcomings we noted here.
In all of the tests we conducted the GAN showed two modes of failure. First it failed at replicating the tails of a distribution, and second, it failed when the distribution to be simulated became more complex; i.e. multiple peaks, or sharp transitions. 
Beyond simply searching for a better architecture or increasing the amount of training data we note three ideas for further study that may lead to better results. The first is to explore the effect of the distribution used to sample the latent space. 
Since the statistics we wish to recover have skewed distributions it may be beneficial to sample the latent space with a skewed distribution. 
Second, it might be helpful to include some sensitivity to tails in our stopping criteria. The criterion we used for training is only sensitive to one statistic and is insensitive to the tails of said distribution as it only relies on the mean and covariance of the distribution of the power spectrum. Finally, it might also be helpful to limit the training set to the level of foreground emission closer to that expected in the survey under consideration. Decreasing the dynamic range of contamination will lessen the challenge of modeling the tails.

Next we trained a Bayesian ResUNet to recover the CMB intensity signal from maps contaminated with galactic dust emission. The training set was created from a combination of CMB realizations and samples from our GAN. Our testing and validation data sets were created from CMB realizations contaminated with the training data we used for the GAN, measurements of interstellar dust intensity by $\planck$.
The training set proved to be robust enough to produce precise estimates of the underlying CMB signal on the test set. 

We chose to approach the task of foreground removal with a Bayesian ResUNet as this type of network learns a distribution over possible functions to clean the input data and we obtained estimates from this distribution through Monte Carlo. 
When compared to a Frequentist approach of estimating uncertainties we found the confidence intervals produced by the ResUNet were overly conservative and sensitive to the dropout rate of each Dropout layer. 
We believe this technique may prove useful for cleaning foregrounds in upcoming analyses and leave
the problem of fine tuning the dropout rates to produce less conservative error estimates for future work.

In both the GAN and the ResUNet we were able to obtain better performance by having the network act on both the map and power spectrum levels. 
In the GAN this was accomplished by adding an extra discriminator and in the the ResUNet we had the network estimate both the map and power spectrum of the CMB. 
These results suggest further improvements could be gained by having the networks act on even higher order statistics, such as the trispectrum.
 
Ultimately while there is room for improvement in both parts of this analysis our work has been successful enough to warrant further exploration and expansion to polarization. 
Expanding this work may prove useful in future experiments where a precise cleaning of foregrounds and an understanding of the distribution of foreground residuals will be necessary for detection of primordial gravitational waves. 
We believe there are many more problems in cosmology that deep learning can provide solutions for and the availability of powerful computation resources makes this route more attractive than ever. 
Deep learning provides a set of tools that may be used to speed up the analysis of data and improve the accuracy of detection of signals of interest and can lead to better and faster constraints on cosmological models.

\acknowledgments
This work used the Extreme Science and Engineering Discovery Environment (XSEDE), that is supported by National Science Foundation grant number ACI-1548562. KA was supported in part by NSF award PLR-1248097. LK was supported in part by NSF awards 1836010 and DMS-1812199.

\bibliography{main}
\end{document}